\documentclass[twocolumn,showpacs,preprintnumbers,amsmath,amssymb]{revtex4}

\usepackage{amsmath,amssymb}
\usepackage{graphicx}
\usepackage{dcolumn}
\usepackage{bm}

\usepackage{graphicx,color}

\def\Beq{\begin{equation}}
\def\Eeq{\end{equation}}

\def\Bea{\begin{eqnarray}}
\def\Eea{\end{eqnarray}}

\def\Beaa{\begin{eqnarray*}}
\def\Eeaa{\end{eqnarray*}}

\def\BD{\begin{description}}
\def\ED{\end{description}}

\def\BC{\begin{center}}
\def\EC{\end{center}}

\def\Bcenter{\begin{center}}
\def\Ecenter{\end{center}}

\def\<{\langle}
\def\>{\rangle}
\def\({\left(}
\def\){\right)}
\def\->{\rightarrow}
\def\-->{\longrightarrow}

\def\b{\beta}

\def\D{\Delta}

\def\l{\lambda}

\def\t{\tau}
\def\s{\sigma}

\def\^{\wedge}

\input{epsf.tex}

\begin{document}

\title{Switching effect upon the quantum Brownian motion near a reflecting boundary}

\author{Masafumi Seriu}
 \email{mseriu@edu00.f-edu.fukui-u.ac.jp}
 \affiliation{%
 Department of Physics, Graduate School of Engineering, 
University of Fukui, 
Fukui 910-8507, Japan\\}%

\author{Chun-Hsien Wu}%
 \email{chunwu@phys.sinica.edu.tw}
\affiliation{%
Institute of Physics, 
Academia Sinica, Nankang, Taipei 11529, 
Taiwan 
}%



\begin{abstract}
 
The quantum Brownian motion of a charged particle in the electromagnetic vacuum fluctuations is investigated 
 near a perfectly reflecting flat boundary,  
 taking into account the smooth switching process in the measurement.   
 
 Constructing a smooth switching function by 
 gluing together a plateau and the Lorentzian switching tails, 
 it is shown that the switching tails have a great influence on the measurement of 
 the Brownian motion in the quantum vacuum. Indeed, it turns out that 
 the result with a smooth switching function and the one with a sudden switching function 
 are qualitatively quite different. It is also shown that anti-correlations between the switching tails and the 
main measuring part plays an essential role
 in this switching effect. 
 
 The switching function can also be 
  interpreted as a prototype of an non-equilibrium process in a realistic measurement, so that  
the switching effect found here is expected to be significant in actual applications in 
vacuum physics. 
\end{abstract}


\thanks{
One of the author (M.S.) has been supported by the Japan  Ministry of Education, Culture, 
Sports, Science and Technology  with the grant \#14740162.}
\maketitle

\section{\label{sec:1}Introduction}

It is well known that a non-trivial  spectral profile  of 
the vacuum fluctuations produces observable effects. One important example in this category is the 
quantum vacuum near  reflecting boundaries, which   is directly related to 
various applications; 
Casimir effect, quantum effects in the early universe, 
quantum noise in a gravitational-wave detector, and so on.
One way of probing such non-trivial  vacuum fluctuations is to 
study  the Brownian motion of a test particle  released in the vacuum in question \cite{WuFord,YuFord}. 
Another approach is, for instance,  to  investigate the interaction between  a mirror and the nearby  vacuum fluctuations~\cite{JS}.

In the present paper, we study the velocity dispersions of the Brownian motion of a charged test particle in 
the quantized electromagnetic vacuum 
near a  perfectly reflecting, flat boundary. 
(Let us assume that the boundary coincides with  the $x$-$y$ plane ($z=0$) for later convenience.) 
This analysis is along the line of  some preceding studies on the quantum Brownian motion, among which 
the  case of an uncharged, polarizable test particle~\cite{WuFord} and 
the case of a charged test particle~\cite{YuFord} are closely 
related to the present one. 
We here, however, would like to  pay special attention to the influence of 
the switching process in  measuring the velocity dispersions. 

The reason why we focus on the switching process  is as follows. 
In Ref. \cite{YuFord}, Yu and Ford calculated the velocity dispersions of 
  a classical charged particle in the electromagnetic vacuum near a reflecting boundary, assuming 
  a {\it sudden switching} process. 
  Here  the ``sudden switching" process indicates the measurement process 
  in which the detector is abruptly turned on and turned off  at the time $t=0$ and $t=\t$, say, respectively. 
  They reported that the $z$-component of the  velocity dispersion of the test particle, 
  $\langle \D v_{z}^{2} \rangle$, does not  vanish in late time, but shows an asymptotic late-time 
  behavior 
  $\langle \D v_{z}^{2} \rangle \sim C/z^2$ ($C$ is some constant).  
  They interpreted this  behavior as a transient effect due to a sudden switching process.  
  However, a sudden switching may not be very realistic in view of  the uncertainty  principle between 
  time and energy 
  since the sudden switching  implies that  a coupling between the field and the particle is switched on/off 
  instantaneously with some finite energy exchange. 
 Furthermore, the fact that the late-time behavior  of $\langle \D v_{z}^{2} \rangle$ does not depend on 
 the main measuring time $\t$ 
 suggests that there should be  cancellation during the time $\t$.  
If so, the switching tails at the edge of the main measuring process
 might have significant influence on  the result. 
 
 It is desirable, thus,  to reanalyze  the same system  under a more realistic 
   measuring process with smooth switching tails and to see how 
   the late-time behavior of the measured $\langle \D v_{z}^{2} \rangle$ depends on the switching process. 
  The aim of the present paper is to undertake this analysis. 
   
We will see  that, contrary to the macroscopic measurements,  the measurement 
of the quantum vacuum fluctuations is  considerably  influenced by 
 the switching tails in a highly non-trivial manner. In particular 
 the anti-correlation between the main measuring part and the switching tails 
 plays an essential role. 

There are three  time-scales characterizing  the present system as is discussed below. 
We also study  how the results depend on the time scales  
to get basic ideas about when switching is regarded as ``smooth" or ``sudden". 
  
The   quantum switching effect analyzed here is  expected to 
find various applications related to non-stationary aspects 
of the vacuum fluctuations.

In Sec. \ref{sec:review}, we first  review some basic results of the case of sudden switching discussed 
in \cite{YuFord}, and then we show that a smooth switching process do lead to a totally different result.  
In Sec. \ref{sec:Reanalysis}, after introducing  
a reasonable switching function, 
the velocity dispersions of the probe particle are explicitly computed, paying 
special attention to the singular integrals caused by the mirror-reflections of the light signal.
We find that a particular anti-correlation between the main measuring part and the switching tails 
 plays an essential role in the measuring process. 
Sec. \ref{sec:cancellation} is to clarify the origin of the anti-correlation effect 
found in the preceding section and we confirm that 
it indeed comes from the interplay between the measuring part and the switching tails. 
In Sec. \ref{sec:t<2z}, the case in which the measuring time is shorter than 
$2z$ is considered. 
Section \ref{sec:Summary} is devoted for summary and several discussions. 
 
\section{\label{sec:review} Sudden switching and smooth switching}
\subsection{\label{sec:PrecedingStudy} The case of sudden switching}

 Let us first recall the analysis of the sudden switching case  discussed in Ref.\cite{YuFord}. 
 Throughout the paper, the analysis  is  done in  the Minkowski  spacetime with    
 a standard coordinate system $(t,x,y,z)$. 
 
 A flat, infinitely spreading mirror of perfect reflectivity is 
 installed  at $z=0$ and   the quantum vacuum of the electromagnetic field is considered 
 inside  the half space $z>0$.  
 Then we consider the measurement of  the quantum fluctuations of the vacuum by using 
  a classical charged particle with mass $m$ and charge $e$ as a probe. 
  When the velocity of the particle is much smaller than the light-velocity, one can assume that 
  the particle couples solely with the electric field $\vec{E} (\vec{x},t)$. Then 
  the equation of motion for the particle is given by
\Beq
 m  \frac{d\vec{v}}{dt}= e \vec{E}(\vec{x},t)\ \ .
 \label{eq:eqofmotion}
\Eeq
Furthermore,  when the position of the particle does not change 
so much within the time-scale in question, Eq. (\ref{eq:eqofmotion}) is approximately solved to
\Beq
   \vec{v}(\t) \simeq  \frac{e}{m}\  {\int_0}^\t \vec{E}(\vec{x},t) dt\ \ . 
 \label{eq:velocity}
\Eeq
Based on Eq.(\ref{eq:velocity}) along with  $\< E_i (\vec{x} , t)  \>_R=0$,  the 
velocity dispersions of the particle, $\< {\D v_i}^2 \>$ ($i=x, y, z$), are 
given by   
\Beq
\< {\D v_i}^2 \> = 
\frac{e^2}{m^2} \int_0^\t dt' \int_0^\t dt'' 
\< E_i (\vec{x} , t') E_i (\vec{x} , t'')  \>_R 
\label{eq:v2sudden}
\Eeq
with\cite{BroMac}
\Bea
 \< E_z (\vec{x} , t') E_z (\vec{x} , t'')  \>_R 
&=& \frac{1}{\pi^2} \frac{1}{ (T^2 - (2z)^2)^2} 
\label{eq:EzEz} \\
 \< E_x (\vec{x} , t') E_x (\vec{x} , t'')  \>_R 
&=& \< E_y (\vec{x} , t') E_y (\vec{x} , t'')  \>_R \nonumber  \\
&& = - \frac{1}{\pi^2} \frac{T^2 + 4z^2}{ (T^2 - (2z)^2)^3} \ \ , 
\label{eq:ExEy}
\Eea
where   $T:= t'-t''$ and  the suffix ``R" is for ``renormalized". 
(We set $c=\hbar=1$ hereafter throughout the paper.)

 Now the explicit computation of Eq.(\ref{eq:v2sudden}) along with 
 Eqs.(\ref{eq:EzEz}) and (\ref{eq:ExEy}) results in~\cite{YuFord} 
\Bea
\< {\D v_z}^2 \> &\doteq&  
\frac{e^2}{32 \pi^2 m^2} \frac{\t}{z^3}  \ln\left( \frac{2z+\t}{2z-\t}  \right)^2 \ \ , 
\label{eq:DeltaVz}
\\
\< {\D v_x}^2  \> &=& \< {\D v_y}^2  \>  \nonumber \\
  &\doteq&  \frac{e^2}{\pi^2 m^2} 
  \left\{ 
  \frac{\t}{64z^3} \ln\left( \frac{2z+\t}{2z-\t}  \right)^2 \right.  \nonumber \\
 && \qquad \qquad \qquad \left. -\frac{\t^2}{ 8z^2 (\t^2 -4z^2) } 
  \right\} \ \ , 
\label{eq:DeltaVxy} 
\Eea
irrespective of whether $\t > 2z$ or $\t < 2z $. Here we note that 
a regularization using the generalized principal value~\cite{DaviesDavies} would lead to the same result  in Ref.\cite{YuFord}
to get these results when $\t >2z$. 
We introduce a special equality symbol ``$\doteq$" (e.g. in Eqs.(\ref{eq:DeltaVz}) and (\ref{eq:DeltaVxy})) 
and an estimation symbol ``$\approx$" (e.g. in Eqs.(\ref{eq:asymptoticZ}) and (\ref{eq:asymptoticXY}) below)
  to remind us that 
a regularization  should be  
employed to get the result when the integral is  a multi-pole integral. 
Indeed the kernel $\< E_i (\vec{x} , t') E_i (\vec{x} , t'')  \>_R$ possesses 
a double pole and a triple pole for $i=z$ and $i=x,y$, respectively, at $T=2z$. 
Thus a regularization should be employed when $\t >2z$. 

Let us note at this stage that there are two  time-scales characterizing  the present situation.  
One is the measuring time $\t$ and the other is the traveling time $z$ of the light-signal 
from the test particle to the plate.

Now the results Eqs.(\ref{eq:DeltaVz}) and (\ref{eq:DeltaVxy}) 
 yield the asymptotic late-time behavior 
\Bea
&& \< {\D v_z}^2  \> 
\approx \frac{e^2}{4\pi^2 m^2 z^2} + O\left(\left(z/\t\right)^2\right) \ \ ,
\label{eq:asymptoticZ}  \\
&& \< {\D v_x}^2  \> = \< {\D v_y}^2  \>
\approx - \frac{e^2}{3\pi^2 m^2 \t^2} + O\left(\left(z/\t\right)^2\right) \ .
\label{eq:asymptoticXY} 
\Eea
Eq.(\ref{eq:asymptoticZ}) indicates 
 that {\it $\< {\D v_z}^2  \>$ remains finite even in the late-time limit, $ \t/z \--> \infty  $}. 
It would mean that an energy of the order of $\frac{1}{2}m  \< {\D v_z}^2  \>$  
is gained during this process.  

Ref.\cite{YuFord} interpreted this asymptotic behavior of $\< {\D v_z}^2  \>$ in  Eq.(\ref{eq:asymptoticZ}) 
as a transient effect caused by  some  energy change due to  the ``sudden-switching". 
Indeed, the formula Eq.(\ref{eq:v2sudden}) corresponds to the measuring process 
with the sudden-switching in which  the measuring device is 
abruptly switched on and switched off  at the time $0$ and $\t$, respectively.
From the viewpoint of the switching function, this measuring process is 
represented by a  step-function 
\Bea
\Theta (t) 
 &=& 1 \ \ \ \    ({\rm for}\ \ 0 < t <  \t) \nonumber \\
 &=&   0   \ \ \ \  ({\rm otherwise}) \ \ .  
 \label{eq:StepFun} 
\Eea
It consists of the  measuring part of the duration $\t$ and infinitely steep 
switching tails. 
 If   the behavior  could be interpreted as the  transient effect during the switching process, 
 then it is expected to see more or less similar behavior even when
  a different switching process is chosen 
 other  than the sudden switching. Let us study this point next.

\subsection{\label{sec:LorentzianCase} The case of smooth switching}
 
 We now replace  $\Theta (t)$ (Eq.(\ref{eq:StepFun})) with the Lorentzian function 
 as a typical smooth switching function. 
 The Lorentzian  function with the characteristic time-scale $\t$ is
\Beq
f_\t (t) = \frac{1}{\pi} \frac{\t^2}{t^2 +  \t^2 }\ \  , 
\label{eq:Lorentzian}
\Eeq
normalized as 
\[
\int_{-\infty}^\infty f_\t (t) dt  = \t\ \ . 
\]

Instead of Eq.(\ref{eq:v2sudden}), the velocity dispersions shall be given by 
\Bea
&& \< {\D v_i}^2 \> 
\ = \frac{e^2}{m^2} \int_{-\infty}^\infty dt' \int_{-\infty}^\infty dt'' \nonumber \\
&& \qquad \qquad f_\t (t') f_\t (t'') 
\< E_i (\vec{x} , t') E_i (\vec{x} , t'')  \>_R \ \ . 
\label{eq:v2Lorentz}
\Eea
The function $f_\t (t)$ represents solely smooth switching tails without 
any flat measuring part. 
In this case, the model is characterized by  two  time-scales,   
i.e. the switching-duration time $\t$ and the traveling time $z$ of the light-signal 
from the test particle to the plate.

If the asymptotic behavior (Eq.(\ref{eq:asymptoticZ})) 
is  due to the transient effect caused by energy input during the switching process, 
then, a similar kind of behavior  is expected  for Eq.(\ref{eq:v2Lorentz}). 
It turns out, however, these integrals are shown to be
\Bea
\< {\D v_z}^2 \> &\doteq&  
\frac{e^2}{16 \pi^2 m^2 \t^2} \frac{1}{(1+\frac{z^2}{\t^2})^2}  \ \ , 
\label{eq:DeltaVzLorentz}
\\
\< {\D v_x}^2  \> &=& \< {\D v_y}^2  \>  
\doteq  - \frac{e^2}{16 \pi^2 m^2 \t^2} 
\frac{ 1- \frac{z^2}{\t^2} }{(1+\frac{z^2}{\t^2})^3}  \ \ . 
\label{eq:DeltaVxyLorentz} 
\Eea
Thus we see that 
\begin{description}
\item{(i)} The short time behavior of $\< {\D v_i}^2  \>$  ($\t \ll 2z$) is same;
\Bea
 \< {\D v_z}^2  \> &\sim&  \left(\< {\D v_x}^2  \> =  \< {\D v_y}^2  \>\right) \nonumber \\
& \sim & 
\frac{e^2}{16 \pi^2 m^2 z^2}\frac{\t^2}{z^2} + O(\left(\t/z\right)^4)\ \ , 
\label{eq:ShortAsympt} 
\Eea
for both the step-function case and the Lorentzian switching case.  
\item{(ii)} However, the long time behavior ($\t \gg 2z$) of the $z$ component 
is quite different. 
For the Lorentzian switching case, it turns out that 
\Bea
&& \< {\D v_z}^2  \> \approx 
\frac{e^2}{16 \pi^2 m^2 \t^2}  
+ O(\left(z/\t \right)^4) 
\label{eq:LorentzAsympt}  \\
&& \< {\D v_x}^2  \> = \< {\D v_y}^2  \>  \approx 
- \frac{e^2}{16 \pi^2 m^2 \t^2}  + O(\left(z/\t\right)^4) 
\label{eq:LorentzAsympt2}
\Eea
\end{description}
Thus  the late-time behavior of 
$\< {\D v_z}^2  \>$ in the Lorentzian switching  case 
is quite different from the step-function case (Eq.(\ref{eq:asymptoticZ})); 
as $\t /z \rightarrow \infty$, the former  goes away 
 while the latter remains finite independently of  $\t$.
The qualitatively different late-time  behavior of $\< {\D v_z}^2  \>$ 
shown in Eq. (\ref{eq:asymptoticZ}) and Eq. (\ref{eq:LorentzAsympt}) is quite puzzling. 
The former depends on $z$, an intrinsic scale of the system, and remains even  in the 
late time, while the latter does not depend on $z$ and goes away in the late time.

It can be said that both the switching functions  studied so far are not  realistic enough. 
On the one hand, a sudden switching could  have virtually picked up the contribution from 
the highly fluctuating vacuum, which  might have been forbidden by the uncertainty principle. On the other hand, 
the pure Lorentzian switching model we have just investigated lacks a plateau of the measuring part, which 
is not realistic either: 
In a proper measurement, the measuring time scale $\t$ should be large enough compared to $z$, the intrinsic 
scale of the system, so that the measuring function is regarded as nearly flat  except for the switching ends.

 We shall  introduce a switching function which blends smoothly  
the step-function and 
the Lorentzian-tails of arbitrary duration in the arbitrary ratio.  In the next few sections, we shall 
construct such a generalized model and investigate it 
in detail.

\section{\label{sec:Reanalysis} Analysis of the velocity fluctuations  of a probe particle with 
a Lorentz-plateau switching function}

Here we undertake the reanalysis of the model introduced above with a more realistic switching function.
We start with  constructing a reasonable switching function. 

\subsection{\label{sec:LPfunction} Lorentz-plateau switching function}

We here  construct a switching function  
which is characterized by a stable  measuring-part (of a time-scale $\t_1$) and two switching-tails describing 
the turn-on and the turn-off processes (of a total time-scale $\t_2$). 
This ``Lorentz-plateau" function $F_{\t \mu} (t)$ is a blend of a plateau part  and the Lorentz function,  
 characterized by  two parameters $\t$ and $\mu$ and  
defined as 
\Bea
F_{\t \mu} (t) 
 &=& 1 \qquad \qquad  \qquad \qquad \ \ \   ({\rm for}\ \ |t| \leq \t/2) \nonumber \\
 &=&   \frac{\mu^2}{ (|t|/\t- 1/2)^2+ \mu^2}  \ \  ({\rm for}\ \ |t| > \t/2).  
 \label{eq:Lplat} 
\Eea
 Its flat part (corresponding to the main measuring period) is matched to  two  tail-parts 
 (corresponding to switching tails) 
  each of which is  half of the Lorentzian function. 
 The smoothness of the matching ($C^1$-class) is  enough for our 
 analysis  since  partial-integrals are not included in our analysis scheme. 
 (If one wishes, however, one may anytime modify $F_{\t \mu}$ to a smoother one.)
 The time-scale characterizing the  measuring-part is $\t:=\t_1$, while 
 $\t_2:=\pi \mu \t$ characterizes the time-scale of the switching-tails: 
\Bea
&& \int_{-\t/2}^{\t/2} F_{\t \mu} (t) dt = \t =\t_1 \ \ , \ \ \nonumber \\
&& 2 \int_{\t/2}^{\infty} F_{\t \mu} (t) dt = \pi \mu \t =:\t_2\ \ . 
\label{eq:tau1tau2}
\Eea
Thus the parameter 
\Beq
\mu= \frac{\t_2}{\pi \t_1} 
\label{eq:mu}
\Eeq
is the switching-duration parameter which  
characterizes  the switching duration relative to the main measuring time-scale. 
In this general setting,  
the situation  described by Eq.(\ref{eq:v2sudden}) 
(as considered in  Ref.\cite{YuFord})  corresponds to the limit 
$\mu \--> 0$ with a fixed $\t_1$. 
Let us call this limiting situation the {\it sudden-switching} 
limit, for brevity. On the other hand, 
the situation  described by 
 Eq.(\ref{eq:v2Lorentz})   corresponds to the limit $\mu \--> \infty$ with a fixed $\t_2$ 
(i.e. $\t_1 \rightarrow 0$ with a fixed $\t_2$), which shall be called the {\it Lorentzian} limit.  
In  most part of the analysis below, it suffices to assume $\mu$  to be less than 1. 
(However, the case $\mu \gg 1$ is also considered when necessary.) 

\begin{figure}[t]
\begin{center}
\includegraphics[scale=0.8]{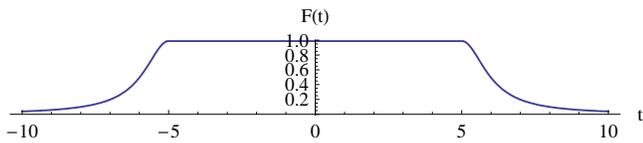}
\caption{Typical example of the Lorentz-plateau function.}
\end{center}
\end{figure}


In view of Eqs.(\ref{eq:v2sudden}) and (\ref{eq:v2Lorentz}) along with Eqs.(\ref{eq:EzEz}) and (\ref{eq:ExEy}),   
what we need to estimate is  of  the form 
\Beq
{\cal I} = \int_{-\infty}^\infty dt' \int_{-\infty}^\infty dt''\  F_{\t \mu} (t') F_{\t \mu} (t'') {\cal K} (t'-t'') \ \ ,   
\label{eq:intbasic}
\Eeq
where ${\cal K}$ is an even  function of $T:=t'-t''$ with an appropriate asymptotic  behavior 
as $|T| \rightarrow \infty$. General properties of the integral Eq.(\ref{eq:intbasic}) are 
analyzed in {\it Appendix} \ref{app:0}. 

Introducing dimension-free variables  $x:=(t'-t'')/\t$ and $y:=(t'+t'')/\t$, 
Eq.(\ref{eq:intbasic}) becomes 
\Bea
{\cal I} &=& \frac{\t^2}{2}\int_{-\infty}^\infty dx \int_{-\infty}^\infty dy \nonumber \\
&& \ \ \ 
    F_{\t \mu} \Big(\frac{\t}{2}(x+y) \Big) F_{\t \mu} \Big(\frac{\t}{2}(y-x)\Big)\  {\cal K} (\t x) \ . 
\label{eq:intbasic2}
\Eea
Now the kernel ${\cal K}$ is essentially 
a  two-point time-correlation function, so that the integral region for  Eq.(\ref{eq:intbasic2}) 
 is naturally divided into 
 4 classes of sub-regions,  $M$,  $S_1$, $S_2$ and $MS$ 
(``$M$" and ``$S$" are  for ``measuring" and ``switching", respectively). Namely 
the class $M$ comes from the two-point correlation solely within the measuring part ($|t|< \t/2$), 
the class $S_1$  from the one within the same switching tail (either 
$t > \t/2$ or $t < - \t/2$),  the class $S_2$ from the one between different  switching tails 
($t > \t/2$ and $t < - \t/2$) and finally 
the class $MS$ from the two-point correlation between the measuring part ($|t|< \t/2$) and the switching 
tails ($|t| > \t/2$). (See {\it Appendix} \ref{app:0} for computational details.)

\subsection{\label{sec:VelDis} Estimation of velocity dispersions using the Lorentz-plateau switching function}

Having prepared a reasonable switching function, 
we now estimate the velocity dispersions of a probe-particle
near a perfectly reflecting plate. 

First let us note that, the model to be analyzed  is now characterized by  three  time-scales 
rather than two,  
i.e. the measuring time $\t=\t_1$,  the switching-duration time $\t_2$ and the traveling time $z$ of the light-signal 
from the test particle to the plate.

Let us focus on $\< {\D v_z}^2 \> $. 
We can make use of  general formulas given in {\it Appendix} \ref{app:0}. 
Comparing  Eq.(\ref{eq:EzEz}) with Eq.(\ref{eq:intbasic}), we can set  
\Beaa
{\cal K}(T) :&=& \frac{e^2}{\pi^2 m^2} \frac{1}{(T^2 -(2z)^2)^2} 
\label{eq:kernel} \\
                &=& \frac{e^2}{\pi^2 m^2} \frac{1}{\t_1^4} \frac{1}{(x^2 -\s_1^2)^2} 
                = \frac{e^2}{\pi^2 m^2} \frac{1}{\mu ^4 \t_1^4} \frac{1}{(\chi^2 -\s_2^2)^2} \ \ . 
\Eeaa
Here several variables and parameters are introduced for simplicity, 
\Bea
&& T:=t'-t''\ ,\  x:=T/\t\ ,\   \chi:= x/\mu\ , \nonumber \\
&&   \s_1:= 2z/\t\ , \ \s_2:= 2z/(\mu \t)=  \s_1/\mu\ \ , 
\label{eq:parameters}
\Eea 
while  $\t=\t_1$ and $\t_2$ are given in Eq.(\ref{eq:tau1tau2}), and 
$\mu$ is given in Eq.(\ref{eq:mu}).

Then Eq.(\ref{eq:intfinal}) in {\it Appendix} \ref{app:0} gives  the formula for $\< {\D v_z}^2 \>$, 
\Bea
&& \< {\D v_z}^2 \> = \frac{2\  e^2}{\pi^2 m^2 \t^2}  \int_0^1 dx \ \frac{1-x}{(x^2 - {\s_1}^2)^2} \nonumber \\ 
&&\ \ \  + \frac{4\     e^2}{\mu^2 \pi m^2 \t^2}
                  \int_0^\infty d\chi\  \frac{1}{(\chi^2+4)(\chi^2 -{\s_2}^2)^2}   \nonumber \\
&&\ \  +  \frac{4\   e^2}{\mu^2 \pi^2 m^2 \t^2}  \int_0^\infty d\chi\ \times      \nonumber \\
&&\ \      \times   \left\{ \frac{1}{(\chi^2 - {\s_2}^2)^2} -\frac{1}{\{(\chi+1/\mu)^2 - {\s_2}^2\}^2} 
             \right\} {\cal F}(\chi)     \nonumber \\
         && =: \< {\D v_z}^2 \>_M + \< {\D v_z}^2 \>_S + \< {\D v_z}^2 \>_{MS} \ \ ,      
\label{eq:intmodel}
\Eea
where ${\cal F}(\chi)$ is given by Eq.(\ref{eq:F-function}).

 Let us investigate three terms  $\< {\D v_z}^2 \>_M$,  $\< {\D v_z}^2 \>_S$ and   $\< {\D v_z}^2 \>_{MS}$
  in Eq.(\ref{eq:intmodel}) in more detail.

We now focus on  the case $\t>2z$ since our main interest  is in  
the late-time behavior of the vacuum fluctuations. (The case $\t < 2z$ shall be treated 
separately  in Sec.\ref{sec:t<2z}.) 
In this case, all of the three integrals in 
Eq.(\ref{eq:intmodel}) are   singular integrals 
since   $0< \s_1 < 1$ and $\s_1 < \s_2 < 1/\mu \leq \infty$.

Let us first estimate the integral $\< {\D v_z}^2 \>_M$, coming from the $M$-region. 
With the help of Eq.(\ref{eq:useful}) in {\it Appendix} \ref{app:2}, 
we get 
\Bea
\< {\D v_z}^2 \>_M:&\doteq& 
  \frac{e^2}{2\pi^2m^2 \t^2}  
 \frac{1}{\s_1^3}\ln\left( \frac{1+\s_1}{1-\s_1} \right) \nonumber \\
 && \sim \frac{e^2}{\pi^2 m^2 \t^2 \s_1^2} \ \ , 
\label{eq:Dv2M}
\Eea
where in the last line $\s_1 \ll 1$ $(\t \gg 2z)$ has been assumed. 
We note that $\< {\D v_z}^2 \>_M$ is the contribution purely from the $M$-region, which 
corresponds to the velocity dispersion  in the case of sudden switching. 
The above expression  exactly coincides with the result given in Ref.\cite{YuFord}.

By a similar prescription for the singular integral along with  
Eq.(\ref{eq:singint1}) in {\it Appendix} \ref{app:1}, we can estimate 
the term $\< {\D v_z}^2 \>_S$, coming  from the $S_1$- and $S_2$-regions, as 
\Bea
\< {\D v_z}^2 \>_S
&\doteq& 
  \frac{\mu^2 e^2}{m^2 \t^2} \frac{1}{(\s_1^2+4\mu^2)^2}  \nonumber \\
&& \sim O\left(\frac{\mu^2}{\s_1^2}\right) \cdot \< {\D v_z}^2 \>_M \ \ ({\rm for}\  \mu < \s_1) \nonumber \\
&& \sim O\left(\frac{\s_1^2}{\mu^2}\right) \cdot \< {\D v_z}^2 \>_M \ \ ({\rm for}\  \mu > \s_1) \ \ .
\label{eq:Dv2S}
\Eea

The term $\< {\D v_z}^2 \>_{MS}$, coming from the $MS$-regions along with the $S_1$- and $S_2$-regions, is estimated as follows. 
Noting that $0 \leq {\cal F} (\chi) <\frac{\pi}{2} $, it follows 
\Beq
 \< {\D v_z}^2 \>_{MS} = O(1) \cdot 
  \frac{2 e^2}{\mu^2 \pi m^2  \t^2}\times   \int_0^{1/\mu} \frac{1}{(\chi^2 - {\s_2}^2)^2} d\chi  \ \ . 
\label{eq:Dv2MSsub}
\Eeq
We note that the integral above is a singular one  since 
$\s_2 < 1/\mu$, which  can be 
treated with the help of Eq.(\ref{eq:singint1}). 
It is notable that, in the above computation for  $\< {\D v_z}^2 \>_{MS}$, the cancellation 
has occurred as is shown the upper-bound of the integral region. Tracing back the origin of this 
cancellation, we can see from the general argument in {\it Appnedix} \ref{app:0}, that 
it comes from the $S_2$-region, which describes the correlation  between 
the pre- and the post-measuring switching tails. 
Thus, this cancellation phenomenon caused by the correlation 
between the pre- and the post-measuring switching tails seems to be quite universal
and  is probably   worth while pursuing further.

The integral can be estimated as 
\Bea
&&\<{\D v_z}^2 \>_{MS} = O(1)\cdot 
  \frac{2 e^2}{\mu^2 \pi m^2  \t^2}\times   \\
\label{eq:Dv2MSsubsub}
&& \times   \left\{ 
              \frac{\mu^2}{2\s_1^2 \rho} + O(\rho) - \frac{\mu^3}{2\s_1^2} 
              \left( \frac{1}{1-\s_1^2} - \frac{1}{2\s_1} \ln\frac{1+\s_1}{1-\s_1}  \right)
             \right\}  . \nonumber 
\Eea
When $\s_1 \ll 1$, it is further modified as 
\Bea
\<{\D v_z}^2 \>_{MS} = O(1) \cdot   \frac{e^2}{\pi m^2   \s_1^2 \t^2} \left(\frac{1}{\rho} - \frac{2\mu \s_1^2}{3}  \right)\ \ .         
\Eea
As mentioned at the end of the previous section, 
the regularization procedure removes the first term on the R.H.S., yielding  
\Bea
\<{\D v_z}^2 \>_{MS}\  & \approx &   - O(1) \cdot \frac{2 \mu e^2}{3\pi m^2  \t^2}  \nonumber \\
                 && \sim - O(\mu \, {\s_1^2}) \<{\D v_z}^2 \>_{M} \ \ . 
\label{eq:Dv2MS}
\Eea
Gathering Eqs.(\ref{eq:Dv2M}), (\ref{eq:Dv2S}) and (\ref{eq:Dv2MS}) together, 
and changing back to the variables $\t_1$,$\t_2$ and $z$, 
we get an estimation for the total velocity dispersion in $z$-direction
\Bea
&&\<{\D v_z}^2\> \nonumber \\
&=&\<{\D v_z}^2\>_M+\<{\D v_z}^2\>_S+\<{\D v_z}^2\>_{MS} \nonumber \\
& \approx & \left\{ 1 +\frac{\pi^4 z^2 \t_2^2}{4(\pi^2 z^2 +\t_2^2)^2}
 - O(1)\cdot \frac{8}{3} \left(\frac{z}{\t_1}\right)^2 \frac{\t_2}{\t_1}  \right\} \<{\D v_z}^2\>_M  \nonumber \\
\Eea
Thus,  under the condition $\t_1 \gg 2z$, we derive the behavior of  the velocity dispersion 
$\<{\D v_z}^2 \>$ as a function of the three parameters $\t_1$, $\t_2$ and $z$: 
\BD
\item{(i)}
When $  \t_2 \ll 2z \ll  \t_1$ , $\<{\D v_z}^2 \> \approx  \<{\D v_z}^2 \>_M$.
\item{(ii)}
When $  \t_2 \approx 2z \ll  \t_1$ ,  $\<{\D v_z}^2 \> \approx \frac{3}{2} \<{\D v_z}^2 \>_M$.
\item{(iii)}
When $  2z \ll \t_1 \ll  \t_2 $ and $\frac{\t_2}{\t_1} = O\left(\left(\frac{\t_1}{2z}\right)^2\right) $, \par
$\<{\D v_z}^2 \>  \approx  \<{\D v_z}^2 \>_{S}$.
\item{(iv)}
When $  2z \ll \t_1 \ll  \t_2 $ and $\frac{\t_2}{\t_1} \gg \frac{\t_1^2}{(2z)^2}   $, \par
$ \<{\D v_z}^2 \>  
        \approx - O\left(\frac{\t_2}{\t_1}\left(\frac{2z}{\t_1}\right)^2\right)\cdot \<{\D v_z}^2 \>_M$  \\
$ \qquad \qquad       \sim  - O(1)\cdot \frac{2\, e^2}{3m^2\pi^2} \frac{\t_2}{\t_1^3}$.
\ED
When the time-scale $\t_2$ of the switching tails is much shorter 
than the  time-scale 
$2z$, the velocity dispersion $\<{\D v_z}^2 \>$ reduces to   the result of the  sudden switching case 
given in Ref.\cite{YuFord} (the case (i)). 
As the 
 time-scale $\t_2$ increases up to  around the time scale $2z$, however, 
 $\<{\D v_z}^2 \>$  becomes  around 3/2 
 times of $\<{\D v_z}^2 \>_M$ (the case (ii)). It means that the contribution from the switching tails, $\<{\D v_z}^2 \>_S$, 
 is almost of the same order as the contribution from the measuring part, $\<{\D v_z}^2 \>_M$.  
 Hence the condition for  the switching to be regarded as 
 the ``sudden switching" is  $\t_2 \ll 2z$, i.e. 
 the switching time-scale is much smaller than the scale characterizing 
 the system configuration. 
 
 Next, as the switching time $\t_2$ increases   
 the velocity dispersion decreases,  reducing to 
 the Lorentzian switching case (Eq. (\ref{eq:Lorentzian})) at around 
 $\t_2 \sim O\left(\left(\frac{\t_1}{2z}\right)^2\right) \t_1$ (the case (iii)).  
 This occurs   mainly due to the cancellation of the $M$-term by the negative contribution from the $MS$-term, 
 which is actually the correlation between the switching part and the main measuring part. 
 
   Finally, the case (iv) shows the possible total negative dispersion 
 when the switching time is really large. However, we should also note that 
 the time-scales cannot be arbitrarily large on account of  the assumption that 
 the position of the particle 
does not  change so much during the whole process of probing the vacuum 
 (see below Eq.(\ref{eq:eqofmotion})). 
 The latter condition can be characterized by 
\Beq
\sqrt{|\<{\D v_z}^2 \>|}\  \Delta T < z \ \ ,
\Eeq
where $\Delta T$ is the time-scale of the whole probing process. 
For the case (iv), we set $\Delta T = \t_2$ to get 
\Beq
\frac{\t_2}{\t_1} < \left(\frac{3\pi^2 m^2 z^2}{2e^2}  \right)^{1/3} \ \ .
\Eeq
To get an idea, let us set $m$ to be the electron mass. Then 
$\frac{\t_2}{\t_1} < 12.7 \left(\frac{z}{\l_e}\right)^{2/3}$ where 
$\l_e$ is the Compton length of the electron ($\sim 10^{-10}$cm). This inequality is likely to be satisfied 
when the system configuration  is so arranged. 
 Here we  point out that this anti-correlation effect can possibly be used to 
control  the total quantum fluctuations in applications.

\section{\label{sec:cancellation} Anti-correlation due to  switching processes}
It has been  found  in the preceding section that 
$\<{\D v_z}^2 \>_{MS}$ becomes  negative after the regularization, which 
plays a key role in the whole  process of vacuum measurement. Since the quantity 
$\<{\D v_z}^2 \>_{MS}$ is the combination of contributions  from the $MS$-, $S_1$- and $S_2$-regions 
(see {\it Appnedix} \ref{app:0}), 
it is desirable to  pin down where the negative correlation comes from. 

Here the  switching function shall be modified 
in three ways to find out  where the negative correlation comes in. 
\subsection{\label{sec:cancellation1} Measuring part with one switching tail}

We choose as a switching function, 
\Bea
F^{(A)}_{\t \mu} (t) 
 &=& 1 \qquad \qquad  \qquad \qquad \ \ \   ({\rm for}\ \ |t| \leq \t/2) \nonumber \\
 &=& 0 \qquad \qquad  \qquad \qquad \ \ \   ({\rm for}\ \ t > \t/2) \nonumber \\
 &=&   \frac{\mu^2}{ (t/\t+ 1/2)^2+ \mu^2}  \ \  ({\rm for}\ \ t < -\t/2).  
 \label{eq:switchingA} 
\Eea
The above switching function is not an even function  in $t$; it consists of  the pre-measurement tail, 
the main measurement part and a sudden switching-off. 

It is easy to see that only the $M$-region and 2 $MS$-regions (among four) contribute to 
the integral Eq.(\ref{eq:intbasic}). 
Thus 
\Bea
&& {\cal I}^{(A)} = {\cal I}^{(M)}+2 {\cal I}^{(MS)} \nonumber \\
&& = 2\t^2 \int_0^1 dx \ (1-x) {\cal K} (\t x)   \nonumber \\
         && \ \ \   +  2 \mu^2 \t^2 \int_0^\infty d\chi\  
             \left\{  
                 {\cal K} ( \mu \t \chi ) - {\cal K} ( \mu \t  (\chi+{1}/{\mu})
             \right\} \tan^{-1}\chi  \ \   \nonumber \\
         &&  \qquad =: {\cal I}^{(A)}_M + {\cal I}^{(A)}_{MS} \ \ .  
\label{eq:intA}
\Eea
 Comparing with Eq.(\ref{eq:intfinal}), it is notable that the behavior of 
 $\tan^{-1}\chi$ in ${\cal I}^{(A)}_{MS}$ is very   similar to 
 ${\cal F}(\chi)$ given in Eq.(\ref{eq:F-function}). Indeed  
 both are 
 monotonically increasing functions which approach $0$ and $\frac{\pi}{2}$ as 
 $\chi$ goes to $0$ and $\infty$, respectively. 
 Thus we notice that ${\cal I}^{(A)}_{MS}$ is qualitatively same  as 
 ${\cal I}_{MS}$ up to  the numerical factor about $\frac{1}{2}$. (The factor around $\frac{1}{2}$ 
 comes because the number of the $MS$-regions is now  2 rather than 4). 
 Afterwards the computations go almost the same as done in Sec.{\ref{sec:Reanalysis}}. 
 Thus we get 
\Bea
 &&  \<{\D v_z}^2 \>^{(A)}\ \  (\sim \<{\D v_z}^2 \>_M + \frac{1}{2}\<{\D v_z}^2 \>_{MS}) \nonumber \\
&&  \approx \frac{e^2}{\pi^2 m^2 \t^2 \s_1^2} 
              - O(1) \cdot \frac{2 \mu e^2}{3\pi m^2  \t^2}  \nonumber \\
                 && \sim \left\{1- O\left(\frac{\mu}{\s_1^2}\right) \right\} \<{\D v_z}^2 \>^{(A)}_{M} \ \ . 
\Eea 
It has turned out that, thus, {\it the negative correlation comes from 
the time-correlation between the measuring part and the switching tail}. 
The result does not change even  when we choose 
$F^{(A')}_{\t \mu}(t):= F^{(A)}_{\t \mu}(-t)$ as a switching function.
It is expected that the switching function $F^{a}_{\t \mu}(t)$ itself is also useful in some applications.

For confirmation, we also consider 2 more modified switching functions below. 

\subsection{\label{sec:cancellation2} Single switching tail}
We choose as a switching function, 
\Bea
F^{(B)}_{\t \mu} (t) 
 &=&   \frac{\mu^2}{ (t/\t+ 1/2)^2+ \mu^2}  \ \  ({\rm for}\ \ t <  -\t/2) \nonumber \\
 &=& 0 \qquad \qquad  \qquad \qquad \ \ \   ({\rm otherwise})  \ \ .
 \label{eq:switchingB} 
\Eea
The above  switching function  consists only of  the half of the Lorentzian function with 
a sudden switching-off. 

Only one $S_1$-region (among the two)  contributes to 
the integral Eq.(\ref{eq:intbasic}). 
Then we get 
\Bea
 \<{\D v_z}^2 \>^{(B)}   & \approx & 
\frac{\mu^2 e^2}{2 m^2 \t^2} \frac{1}{(\s_1^2+4\mu^2)^2} \ \ , \nonumber \\
&\sim& \frac{1}{2} \<{\D v_z}^2 \>_S \ \ .
\Eea
Note that the above result  is half of the result given in Eq.(\ref{eq:Dv2S}). 
There is no change even  when we choose 
$F^{(C')}_{\t \mu}(t):= F^{(C)}_{\t \mu}(-t)$ as a switching function.
Thus, the correlations  within the same switching tail do nothing with the 
negative correlation effect.

\subsection{\label{sec:cancellation3} Switching tails without the measuring part}
We choose as a switching function, 
\Bea
F^{(C)}_{\t \mu} (t) 
 &=& 0 \qquad \qquad  \qquad \qquad \ \ \   ({\rm for}\ \ |t| \leq \t/2) \nonumber \\
 &=&   \frac{\mu^2}{ (|t|/\t- 1/2)^2+ \mu^2}  \ \  ({\rm for}\ \ |t| > \t/2).  
 \label{eq:switchingC} 
\Eea
The above  switching function  consists only of  the switching tails. 
Only the $S_1$- and $S_2$-regions  contribute to 
the integral Eq.(\ref{eq:intbasic}). 

 Following the estimations  in Sec.{\ref{sec:Reanalysis}}, 
 we easily get 
\Beq
  \<{\D v_z}^2 \>^{(C)} 
 \approx \<{\D v_z}^2 \>_S + \frac{4 {\cal Z}  \mu e^2}{3\pi^2  m^2  \t^2} \ \ , 
\Eeq
where the factor ${\cal Z}$ is some numerical factor much smaller than 1 and 
$\s_1 \ll 1$ has been assumed.  
Thus, the correlations between the two switching tails and those within the same switching tail 
do not yield  negative contributions. 
Noting the inequality  $\<{\D v_z}^2 \>^{(C)} > 2 \<{\D v_z}^2 \>^{(B)}$,  
it is seen that {\it the correlations between the two switching tails
cause small enhancement of $\<{\D v_z}^2 \>$}. 
\subsection{\label{sec:cancellation4} Origin of the negative correlation effect}

From the results of the subsections \ref{sec:cancellation1}-\ref{sec:cancellation3}, 
it is now clear that the origin of the negative correlation effect resides in 
the correlation between the measuring part and the switching tail. Furthermore, one 
switching tail along with the measuring part is enough to cause this effect. 
In this way, it is seen that the interplay between the 
measuring part and the switching tail is a key to understand the measurement process 
of quantum vacuum.

\section{\label{sec:t<2z} Analysis for the case $\t < 2z$}

We have mainly studied the case $\t_1 > 2z$ so far. In this section, 
let us investigate the case $\t_1 < 2z$ in some detail.

The analysis goes in the similar manner up to Eq.(\ref{eq:intmodel}). 
We note that $\s_1 > 1$ and 
\[\s_2:=\s_1/\mu  >  {\rm Max}(\s_1, 1/\mu) > {\rm min}(\s_1, 1/\mu) > 1 
\]
in this case. The situation now is that the measuring time-scale $\t_1$ is shorter than 
the intrinsic time-scale $2z$,  and the switching time-scale $\t_2$ is even shorter 
than $\t_1$. 
Contrary to the case $\s_1< 1$ ($\t_1 > 2z$), the first term $\< {\Delta v_z} ^2  \>_M$  in  
 Eq.(\ref{eq:intmodel}), which comes  purely from the $M$-region, 
is now a regular integral due to $\s_1 > 1$ and it exactly matches the original result of the 
step-function case shown in Ref.\cite{YuFord} (Eq.(\ref{eq:ShortAsympt})). 
Physically, it corresponds to 
 the situation in which 
 the  measuring time $\t_1$ is so short that the information exchange between the mirror and 
 the particle has not yet taken place.  Therefore it is  expected that 
 the presence of the reflecting boundary does not play a vital role in this case. This is why  the 
 integral for $\< {\Delta v_z} ^2  \>_M$  is regular when  $\t_1 < 2z$.

On the other hand, $\< {\Delta v_z} ^2  \>_S$, which comes from  the $S_1$- and $S_2$-regions, 
is given by  a singular integral. 
The appearance of the singular integral is understood as the long-tail nature of the 
Lorentzian function.  
Though the measuring-part is too short  to cause  correlations due to the reflecting boundary, 
the long tails of the switching-part  still  pick up 
correlations. 
By a regularization, $\< {\Delta v_z} ^2  \>_S$ is given by  
\Beaa
\< {\Delta v_z} ^2  \>_S & \doteq &
 \frac{\mu^2\  e^2}{m^2 \s_1 ^4 \t^2}\frac{1}{\left( 1+ \frac{4\mu^2}{\s_1^2}  \right)^2} \\
 &\sim & \frac{\mu^2 e^2}{m^2  \s_1^4 \t^2} \ \ .
\Eeaa

Finally  $\< {\Delta v_z} ^2  \>_{MS}$ in Eq.(\ref{eq:intmodel}), coming from 
the $MS$-, $S_1$- and $S_2$-regions,  turns out to be finite. 
By shifting the variable $\chi':=\chi + 1/\mu$ in the second term, it can be estimated as 
\[
\< {\Delta v_z} ^2  \>_{MS} \sim O(1) \cdot \frac{2 e^2}{\pi m^2 \mu^2  \t^2}\int_0^{1/\mu} \frac{1}{(\chi^2 -\s_2^2)^2} d\chi\ \ , 
\]
which is a regular integral since $\s_2 > 1/\mu$. 
Performing the integral, we get 
\Beaa
 \< {\Delta v_z} ^2  \>_{MS} & \sim &  
O(1)\cdot  \frac{\mu\  e^2}{2 \pi m^2 \s_1^3  \t^2} \ln \left( \frac{\s_1 +1}{\s_1-1} \right) \\
     && \qquad  +O(1) \cdot  \frac{\mu\  e^2}{ \pi m^2 \s_1^4  \t^2}  \frac{1}{1-\frac{1}{\s_1^2}} \\
     &\sim &  O(1)\cdot  \frac{2 \mu\  e^2}{ \pi m^2  \s_1^4  \t^2 } \ \ .
\Eeaa
It is curious  that $\< {\Delta v_z} ^2  \>_{MS}$ does not contain any singular 
integral when $\t <2z$ although the Lorentzian switching tails take part in $\< {\Delta v_z} ^2  \>_{MS}$.
 The reason may be that 
this sector contains the description of   cancellations between the pre- and post-measurement tails (the $S_2$ region). 

 Leaving only  the most dominant terms, we get the estimation 
\Bea
 \< {\Delta v_z} ^2 \> 
& \approx & \frac{e^2}{\pi^2 m^2 \s_1^4  \t^2}\ \ \    ({\rm for}\  \mu \s_1 < 1) \ \ ,
 \label{eq:Dv2zearly} \\ 
& \approx &  O(1)\cdot 
\frac{2 \mu\ e^2}{ \pi m^2  \s_1^4 \t^2}  \ \  ({\rm for}\  \mu \s_1 >1) \ \ ,  
 \label{eq:Dv2zearly2}
\Eea
where Eq.(\ref{eq:Dv2zearly}) comes from the $M$-region and coincides with 
Eq.(\ref{eq:ShortAsympt}), while Eq.(\ref{eq:Dv2zearly2}) comes from 
the $S_1$-,  $S_2$- and $MS$-regions.

\section{\label{sec:Summary} Summary and discussions}

 In the present paper, the effect of a switching process  upon the measurement of 
 the Brownian motion of a charged test particle  
 near a perfect reflecting boundary has been investigated. 
 
We have started with the fact that the late-time asymptotic behavior of 
$\< {\Delta v_z} ^2  \>$ does not depend on the measuring time $\t_1$ but only on the distance to the boundary, $z$, 
when the sudden-switching is employed (Eq.(\ref{eq:asymptoticZ}))~\cite{YuFord}. 
 The $\t_1$-independence suggests that effective cancellations should be taking place 
 during the measuring process and, if so, 
 the result  should be sensitive to the switching-tails at the edge of the measuring part.
This  is a natural reasoning considering the highly fluctuating nature of the vacuum. 
In the measurement of a  normal system,    the switching effect is likely to be ignored 
if the measuring time-scale is much larger than  the switching time-scale. 
When dealing with the quantum vacuum like the present case, however,  the highly fluctuating 
 vacuum might cause 
 the cancellations during the main measuring process so that 
the switching effect can also be an important ingredient.  

Next we have proceeded to calculate 
the velocity dispersion with a Lorentzian function, which represents a pure smooth switching-process 
without a flat measuring part. We have then shown that 
the result is very different from the sudden-switching case (Eq.(\ref{eq:LorentzAsympt})).

Finally constructing the  Lorentz-plateau switching function, 
we have shown that  the result is very sensitive to the switching tails in probing the quantum vacuum. 
We have also derived a reasonable criteria for the switching to be regarded as ``sudden" or ``smooth".
Only the condition that  the switching time-scale $\t_2$ is much smaller than the measuring time-scale $\t_1$
is not enough for the switching to be  qualified as ``sudden". 
As clarified in  \ref{sec:Reanalysis} (the case (i) there),  
the criteria for the validity of the sudden-switching  approximation should be $\t_2 \ll z$ as well as 
$\t_2 \ll \t_1$, where $z$ is interpreted as the traveling time of a signal 
from the particle to the reflecting boundary.

The above  criteria, however,  may not be easy to be satisfied  so that 
the sudden-switching approximation should be taken care more carefully when we 
consider  an actual procedure of measurement. 
We can imagine  an  example for measuring 
$\< {\Delta v_z} ^2  \>$ as follows. 
Assume a wide conducting  plate of a square shape (the edge size $L$) is fixed in the vacuum. 
For clarity of the argument, let us introduce the standard $(x, y, z)$-coordinates in such a way as the 
plate is  contained in the $x$-$y$ plane with the original point O being at the center of the plate.  
From a distant point $P (-A, 0, z)$ ($A \gg L$ and $z \ll L$), 
 a charged particle is shot parallel to the plate with an incident
velocity $\vec{v}_0=(v_0, 0, 0)$.  
The particle  initially goes in  the empty space far away from the plate and then passes near 
the edge of the plate, and finally  enters  into the region bounded by  the  plate. 
The smooth switching function discussed in the paper
 would be   interpreted    as a mathematical description of this process of shooting a probe from a far distance.
In this situation, the switching time scale is 
around $z/v_0$ and the intrinsic time-scale determined by the system configuration is $z$. 
It is obvious  that the switching time-scale can not be smaller than the intrinsic time scale $z$ 
in this case, since $v_0 < 1$.
 This is just one example, but it at least   shows that the sudden-switching approximation is not valid all the time
  and that we should be more careful about the switching effect in dealing with the quantum vacuum. 

In view of the above example, it might also be  possible to look at  the switching function from a different angle, i.e. 
 as a mathematical description of what the test particle would experience 
 when  the vacuum shifts from the  Minkowski vacuum to a Casimir-like vacuum. Based on this interpretation, 
 it is not surprising to see $\< {\Delta v_z} ^2  \>$ 
remain  constant in the late-time, depending only on $z$ for the sudden switching case. 
For, it is interpreted as  the sudden energy shift due to the sudden change of vacuum state.  
In the  case of the pure Lorentzian switching-function, on the other hand, 
the corresponding interpretation is  that the vacuum changes smoothly  from the asymptotic Minkowski vacuum to 
the Casimir-like vacuum, going back to the asymptotic Minkowski vacuum again. Then 
the test particle is never stabilized in this varying vacuum so 
that the result is naturally so different from the sudden switching case.  
Then the setup using Lorentz-Plateau function in this connection  would be interpreted as 
describing a smooth transient process from the Minkowski vacuum to the Casimir-like vacuum.
Thus it is expected that the Lorentz-Plateau function constructed in the present paper might 
be very useful to analyze the situations such as a smooth transient from one vacuum to another. 
 
 Finally it is appropriate to make some comments on the 
 singular integrals and their regularization procedure. 
Tracing back the origin of the singular integral, it  comes from 
the singularity at $T=2z$ in  the integral kernel (Eq.(\ref{eq:kernel}) or Eq.(\ref{eq:EzEz})). 
This singularity is understood as produced  by the reflecting boundary.   
Due to the mirror-reflections of 
signals with the light-velocity, the values of the electric field at 
the two world-points $(t', x, y, z)$ and $(t'' , x, y, z)$ 
are expected to be strongly correlated when $T=t' -t''=2z$. 
These correlations accumulate in  the velocity fluctuations of the particle at $z$ 
when the measurement time $\t$ is longer than the travel time $2z$ for the signals. 
It is natural, thus, to expect that 
the resulting  singular terms of the form $A/\rho$ ($A>0$ and $\rho \--> 0$) 
contain the information on the reflecting boundary. 
However the standard regularization procedure\cite{DaviesDavies} corresponds to discarding  
these singular terms in effect. 
It should be clarified when this type of 
regularization is valid and when not. 

With the above physical interpretation of the singular integrals, 
another natural  way of regularization should be possible. It has been assumed that 
the probe particle and the reflecting boundary or the  mirror are treated as classical objects. However in reality 
they also cannot escape quantum fluctuations.  Taking into account their quantum fluctuations, 
the effective path-lengths of signals are expected to be vague. It is estimated that 
the quantum fluctuations of the probe particle are more significant than those of the mirror. 
It is natural to assume the effective size of the particle to be of the order of its 
Compton length $\l_c= 1/m$, corresponding to setting the infinitesimal parameter $\rho$ to be 
$\rho = \l_c /\t =\frac {1}{m\t} $.
Just for an illustration, let us consider the case of an electron ($\l_c \sim 10^{-10}$ cm) with 
 $\t=1 $ $\mu$sec. Then  $\rho \sim 10^{-15}$. 
It turns out that, thus,  
the singular terms would  all the time dominate in the velocity fluctuations. 
Since the results could be drastically influenced, 
it should also be clarified whether  the cut-off type  of 
regularization is valid. 


\section*{Akcnowledgement}
We would like to thank L. H. Ford for various helpful comments 
on many aspects on the present research.

\appendix

\section{\label{app:0} General features of the integral with a  Lorentz-plateau switching function}

We here analyze general properties of   the integral 
given in Eq.(\ref{eq:intbasic}) or Eq.(\ref{eq:intbasic2}).

We note that  the $x$-$y$ plane  is 
divided into 9 integral regions by 4 border lines, $x+y=\pm 1$ and 
$y-x=\pm 1$, and  the 9 regions are further  classified into  
4 classes,  $M$,  $S_1$, $S_2$ and  $MS$, as discussed after Eq.(\ref{eq:intbasic2}).
In each integral region, the $y$-integral  can be 
done independently of the kernel ${\cal K}$, leaving  the $x$-integral.  
Now we shall investigate each of 4 types of integral regions one by one. 

\begin{figure}[t]
\begin{center}
\includegraphics[scale=0.4]{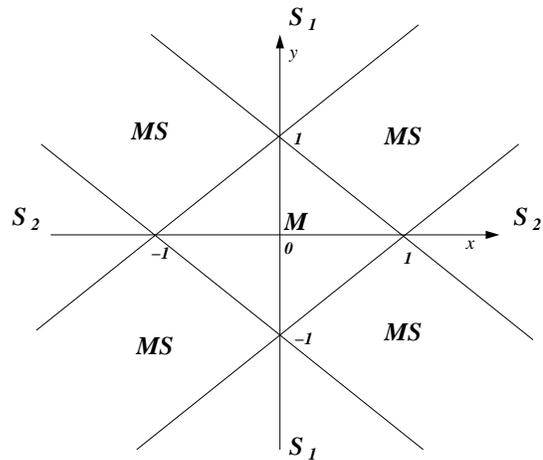}
\caption{Illustraion of four types of integral regions.}
\end{center}
\end{figure}

\BD
\item{{\it (i) $M$-Region :}} {\it The region defined by $|x+y|\leq 1$ and $|x-y| \leq 1$.} \\
It coincides with  the sudden-switching case  considered in Ref.\cite{YuFord}. 
The integral ${\cal I}^{(M)}$, coming from this region,  is computed as 
\Bea
&& {\cal I}^{(M)} \nonumber \\
&&\ =  \frac{\t^2}{2} \left(\int_{-1}^0 dx \int_{-x-1}^{x+1} dy + \int_0^1 dx \int_{x-1}^{-x+1} dy   \right) 
        {\cal K} (\t x)  \nonumber \\
&& \  =  2\t^2 \int_0^1 dx\  (1-x) {\cal K} (\t x) \ \ , 
\label{eq:intM}
\Eea
where the last line follows using the even function property of ${\cal K}$. 
 \item{{\it (ii) $MS$-Regions :}} 
{\it The 4 regions defined by `$x+y \geq 1$ and $|y-x| \leq 1$', 
`$|x+y| \leq 1$ and $y-x \geq 1$, 
`$x+y \leq -1$ and $|y-x| \leq 1$,  and 
  `$|x+y| \leq 1$ and $ y-x \leq -1$'.} \\

For illustration, let us focus on   the region `$x+y \geq 1$ and $|y-x| \leq 1$'.   
The integral ${\cal I}^{(MS)}$, coming from this region,  is computed as  
\Bea
&& {\cal I}^{(MS)} \nonumber \\
&&\ =  \frac{\t^2}{2}  \left(\int_{0}^1 dx \int_{-x+1}^{x+1} dy 
       + \int_1^\infty dx \int_{x-1}^{x+1} dy   \right) \cdot  \nonumber \\
&& \qquad  \qquad \qquad      \cdot   \frac{\mu^2}{\left( \frac{x+y}{2} - \frac{1}{2}  \right)^2 + \mu^2}{\cal K} (\t x)
                                                              \nonumber   \\
&& =   4\t^2 \mu \int_0^\infty dx\ \left(  {\cal K} (\t x) - {\cal K} (\t (x+1)\right) \times \nonumber \\
                       && \qquad \qquad  \times         \tan^{-1} \frac{x}{\mu}   \ \ , 
\label{eq:intMS}              
\Eea      
where the even function property of ${\cal K}$ has been employed to get the  result.
It turns out that each of the 4 regions yield  exactly the same contribution ${\cal I}^{(MS)}$ given 
by Eq.(\ref{eq:intMS}). 
\item{{\it (iii) $S_1$-Regions :}}  {\it  The 2 regions defined by `$x+y \geq 1$ and $y-x \geq 1$' and 
  `$x+y \leq -1$ and $y-x \leq -1$'.} \\
By performing the $y$-integral  and using the even function property of ${\cal K}$,  
it turns out that these 2 regions yield  the same contribution,   
\Bea
&& {\cal I}^{(S_1)}
 = 2 \t^2 \mu^3 \int_0^\infty dx\ 
\frac{{\cal K} (\t x)}{x^2 + 4 \mu^2} \times \nonumber \\
 && \ \ \ \times \left\{ \pi -  \tan^{-1} \frac{x}{\mu} 
-  \frac{\mu}{x} \ln \left( 1+ \frac{x^2}{\mu^2} \right)    \right\} .
\label{eq:intS_1}      
\Eea
 \item{{\it (iv) $S_2$-Regions :}} {\it The two regions defined by `$x+y \geq 1$ and $y-x \leq -1$' and
 `$x+y \leq -1$ and $y-x \geq 1$'.} \\
By performing the $y$-integral  and using the even function property of ${\cal K}$,  
it turns out that these 2 regions yield  exactly the same contribution,   
\Bea
{\cal I}^{(S_2)}&=&  2 \t^2 \mu^3 \int_0^\infty dx\ 
\frac{{\cal K} (\t (x+1))}{x^2 + 4 \mu^2} \times \nonumber \\
&& \times \left\{  \tan^{-1} \frac{x}{\mu} + 
\frac{\mu}{x} \ln \left( 1+ \frac{x^2}{\mu^2} \right)    \right\}  . 
\label{eq:intS_1prime}      
\Eea
\ED

Gathering the results Eqs.(\ref{eq:intM})-(\ref{eq:intS_1prime}), we get 
\Bea
 &&\qquad  {\cal I}= {\cal I}^{(M)}+4{\cal I}^{(MS)}+2{\cal I}^{(S_1)}+2{\cal I}^{(S_2)} \nonumber \\
        && = 2\t^2 \int_0^1 dx \ (1-x) {\cal K} (\t x) 
          +   4 \pi  \mu^2 \t^2 \int_0^\infty d\chi\  \frac{{\cal K}(\mu \t \chi)}{\chi^2 + 4 }   \nonumber \\
         && \ \ \   +  4 \mu^2 \t^2 \int_0^\infty d\chi\  
             \left\{  
                 {\cal K} ( \mu \t \chi ) - {\cal K} ( \mu \t  (\chi+{1}/{\mu})
             \right\} {\cal F} (\chi)  \ \   \nonumber \\
         &&  \qquad =: {\cal I}_M + {\cal I}_S + {\cal I}_{MS} \ \ ,  
\label{eq:intfinal}
\Eea
with
\Beq
{\cal F}(\chi):= \left(1-   \frac{1}{\chi^2+4}  \right) \tan^{-1}\chi -  \frac{1}{\chi(\chi^2+4)}\ln (1+ \chi^2)\ \ .
\label{eq:F-function}
\Eeq
Here a  variable $\chi:= x/\mu$ has been introduced in ${\cal I}_S$ and ${\cal I}_{MS}$ 
(see Eq.(\ref{eq:parameters}) for definitions of variables and parameters). 

The expression Eq.(\ref{eq:intfinal}) reveals 
several general properties of the integral representation Eq.(\ref{eq:intbasic}).

First of all, the two limiting cases of $\cal I$, $\mu \rightarrow 0$ and $\mu \rightarrow \infty$, 
 can be easily obtained (let us once again 
 recall  $\mu$ is the switching-duration parameter defined in Eq.(\ref{eq:mu})):  
On the one hand, we see  
\Beaa
{\cal I} \longrightarrow && {\cal I}_M=2\t^2 \int_0^1 dx \ (1-x) {\cal K} (\t x) \\
\Eeaa
as $\mu \rightarrow 0$ with  a  fixed $\t$ (or equivalently $\t_2 \rightarrow 0$ with a  fixed $\t$).
This  limiting expression  is a  general formula corresponding  to 
Eq.(\ref{eq:v2sudden}), so that the sudden-switching limit ($\mu \rightarrow 0$ with a fixed $\t$) 
 is well-defined in general. 

On the other hand,  expressing Eq.(\ref{eq:intfinal}) in terms of 
$\t_2$ instead of $\t$, it follows 
\Beaa
{\cal I} \longrightarrow && {\cal I}_S   
                             =   \frac{2\t_2^2}{\pi^2} 
                               \int_{-\infty}^\infty  d\xi \frac{ {\cal K} (\t_2 \xi )}{\xi^2 +4/\pi^2}  \\
\Eeaa
as $\t \rightarrow 0$ with a  fixed $\t_2$ 
 (or equivalently $\t \rightarrow 0$ with  a  fixed $\mu \t$).
This limiting  expression is equivalent to the result obtained from Eq.(\ref{eq:intbasic}) with 
$F_{\t \mu}(t)$ being replaced by the following 
Lorentzian function 
\[
f(t) =\frac{1}{\pi^2} \frac{\t_2^2}{t^2 + (\t_2/\pi)^2} \ \ , 
\]
which is nothing but the limiting function of 
$F_{\t \mu}(t)$ as $\t \rightarrow 0$ with a fixed 
$\mu$ (or equivalently $\t\rightarrow 0$ with a fixed $\t_2$).
Thus the Lorentzian limit is also well-defined in general. 

In this manner, the switching function $F_{\mu \t}(t)$ smoothly bridges 
the gap between 
the step-function and the Lorentz-function and 
is expected to be quite useful  for investigating 
various switching effects in quantum vacuum.

Next, let us assume ${\cal K}(T) > 0$ for the sake of  later application. 
It is easy to see that ${\cal F}(\chi)$ is a monotonically increasing function 
with ${\cal F}(0)=0$ and $\lim_{\chi\rightarrow \infty}{\cal F}(\chi) =\frac{\pi}{2}$. 
Then one can estimate 
\[
{\cal I}_{MS} \sim O(1)\cdot 2 \pi \mu^2 \t^2  
                \int_0^{1/\mu} d\chi\  {\cal K} ( \mu \t \chi )  \ \ .
\]
Since ${\cal I}_M$ and ${\cal I}_{MS}$ are estimated by integrals with 
a compact integral region, they may or may not be singular integrals depending 
on whether the pole of ${\cal K}(T)$ is included inside their integral regions. 
In our model, the case  $\t > 2z$ makes them singular while the case $\t < 2z$ non-singular. 
On the other hand, ${\cal I}_S$ is always a singular integral due to the long-tail nature 
of the Lorentzian function. 
When integrals are singular due to the pole of the kernel ${\cal K} (T)$, 
some regularization procedure can enter the analysis. 
The  regularization employed in the present context\cite{DaviesDavies} (following Ref.\cite{YuFord}) 
is in effect to throw away 
an infinitely large positive terms. (See arguments after Eq.(\ref{eq:DeltaVxy}).) 
Thus the apparent positive quantity can become negative after regularization.
We encounter this situation in Sec. \ref{sec:VelDis}.  
In the  case discussed in Sec. \ref{sec:VelDis}, the integrals ${\cal I}_M$ and ${\cal I}_S$ remain positive, while 
${\cal I}_{MS}$ becomes negative after regularization. What happens then is that, 
as the switching time-scale becomes longer,  
either ${\cal I}_S$ or ${\cal I}_{MS}$ dominates depending on the choice of the time-scale parameters.
When ${\cal I}_S$ dominates, the situation is close to 
the Lorentzian smearing case (Eqs.(\ref{eq:Lorentzian})-(\ref{eq:LorentzAsympt})). 
 When the negative term ${\cal I}_S$ dominates, on the other hand, the velocity dispersion in $z$-direction
 becomes negative.

\section{\label{app:1} Formulas for  singular integrals}

Here we present the formulas for  particular singular integrals needed in our analysis:
\Bea
{\cal I}(\s, \xi:2) :&=& 
\Re \int_{{\curvearrowleft}_\rho (\s)} \frac{1-\xi z}{(z^2- \s^2)^2} dz \nonumber \\
&=& \frac{1-\xi \s}{ 2\s^2 \rho} + O(\rho)\ \ \ \ (\s, \xi \in {\bf R} ) \ \ , 
\label{eq:singint1} \\
{\cal I} (\s, \xi:3) :&=&
\Re \int_{{\curvearrowleft}_\rho (\s)} \frac{1-\xi z}{(z^2- \s^2)^3} dz \nonumber \\
&=& -\frac{3-\xi \s}{ 8 \s^4 \rho} + O(\rho)\ \ \ \ (\s, \xi \in {\bf R} ) \ . 
\label{eq:singint2} 
\Eea
Here ${\curvearrowleft}_\rho (\s)$  indicates a semicircle (with an anti-clockwise direction) in the 
upper-plane of $z$ with its 
radius being  $\rho$ $(>0)$ and its center located at $z= \s$.   More precisely, 
${{\curvearrowleft}_\rho (\s)}:=\{z \in {\bf C}| z=\s + \rho \: e^{i\theta},\ \theta \in [0, \pi]\}$ with 
its  direction matched with  increasing  $\theta$.

We here derive  only the formula Eq.(\ref{eq:singint1}). One can derive    Eq.(\ref{eq:singint2}) 
in the same manner. 

Now setting $z=\s + \rho \: e^{i \theta}$, it is straightforward  to see that 
\Beq
{\cal I}(\s, \xi:2) = 
- \frac{1}{\rho}  \int_0^\pi \frac{\Im {\cal A}(\theta )}{{\cal D}^2(\theta)}\  d\theta \ \ ,
\label{eq:int1}
\Eeq
where ${\cal A}(\theta ):=(2\s + \rho e^{-i\theta})^2 ((1- \xi \s) e^{-i\theta}  -\xi \rho)$ and 
${\cal D}(\theta):= 4\s^2 +\rho^2 + 4\s \rho \cos \theta$.  

Due to the relation, $\cos\theta = \{{\cal D}(\theta ) -(4\s^2 +\rho^2)\} / 4\s\rho$, the imaginary part of 
the function ${\cal A}(\theta )$ can be expressed in  powers of $\cal D$:
\Beaa
\Im {\cal A}(\theta ) 
=  - (p\: {\cal D}(\theta)^2   - q\:  {\cal D}(\theta)  + r ) \sin\theta\ \ , 
\Eeaa
where 
\Beaa
&& p:=\frac{\b}{4\s^2}\ \ , \ \ q:= \frac{\rho^2}{2\s^2} \ \ , \\
&& r:= \b \left(\frac{\rho^4}{4\s^2} - \rho^2 \right) + \xi \left( \frac{\rho^4}{2\s} -2\s \rho^2 \right)\ \ , 
\Eeaa
with $\b:= 1-\xi\s$. Thus 
\Bea
&& \rho \ {\cal I}(\s, \xi:2) = \int_0^\pi \left( p - \frac{q}{{\cal D}(\theta)} + \frac{r}{{\cal D}(\theta)^2}   \right) 
               \sin\theta d\theta  \nonumber \\
&& \ \  = 2p -\frac{q}{4\s\rho}\ln \left( \frac{1+\frac{\rho}{2\s}}{1-\frac{\rho}{2\s} }  \right)^2
              + \frac{r}{8\s^4} \left( 1- \frac{\rho^2}{4\s^2}  \right)^{-2} \nonumber \\
&& \ \ = \frac{\b}{2\s^2} + O(\rho^2)\ \ , 
\label{eq:int2}
\Eea
yielding the formula 
Eq.(\ref{eq:singint1}).

We note that the  results remain the same  even when 
${\curvearrowleft}_\rho (\s)$ is replaced by $\{ {{\curvearrowleft}_\rho (\s)} \}^*$ in Eqs.(\ref{eq:singint1}) and 
(\ref{eq:singint2}). Here $\{ {{\curvearrowleft}_\rho (\s)} \}^*$ 
denotes the complex conjugate of the curve ${{\curvearrowleft}_\rho (\s)}$, i.e. 
$\{ {{\curvearrowleft}_\rho (\s)} \}^* :=\{z \in {\bf C}| z=\s + \rho\  e^{-i\theta},\ \theta \in [0, \pi]\}$ 
with its  direction matched with   increasing  $\theta$. 
This is obvious since integrals in  Eqs.(\ref{eq:singint1}) and (\ref{eq:singint2}) are computed from the 
real part of the integrands. 

One can also check the above claim  by the following consideration.
Let $f(z)$ be any function which has an isolated  pole at $z=\s$ with its 
residue, ${\rm Res}(f,\s)$,  being  real. 
Let $C_\rho (\s)$ be a circle (with the anti-clockwise direction) 
of radius $\rho$ with its center  at 
$z=\s$.  When $\rho$ is chosen to be sufficiently small, then,  
\[
\int_{C_\rho (\s)} f = \int_{{\curvearrowleft}_\rho (\s)} - 
\int_{\{{\curvearrowleft}_\rho (\s)\}^*} = 2\pi i\: {\rm Res}(f,\s)\ \ .
\]
Considering the real part of this equation, we get 
\[
\Re \int_{{\curvearrowleft}_\rho (\s)} = \Re \int_{\{{\curvearrowleft}_\rho (\s)\}^*} \ \ .
\]
Thus the claim is confirmed once again by choosing 
$f(z)=\frac{1-\xi z}{(z^2- \s^2)^2}$ or $f(z)=\frac{1-\xi z}{(z^2- \s^2)^3}$. 

\section{\label{app:2} Asymptotic  principal values of singular integrals}

Here we introduce a special treatment for  a singular integral, represented  by 
a symbol $\wp_{(\rho)}$.  Let $f(x)$ be a real function which is   possibly  singular at  $x=\s$.  
Now, for a sufficiently small $\rho$ $(>0)$,  we define 
\Beq
\wp_{(\rho)} \int_A^B\ f(x) dx := \left( \int_A^{\s -\rho} + \int_{\s + \rho}^B \right)\  f(x) dx\ \ , 
\label{eq:pv}
\Eeq
where $A < \s <B$.  
Conventionally, 
if the R.H.S. of Eq.(\ref{eq:pv}) converges as $\rho \rightarrow 0$, 
the value of convergence  is called the principal value of the integral $\int f(x) dx$, denoted 
 by  $p.v. \int f(x) dx$. 
 We here have  generalized  
the concept of the principal value and  have left  the positive small 
quantity $\rho$ as a free parameter. Let us call the above 
$\wp_{(\rho)} \int f(x) dx$  {\it an asymptotic principal value of order $\rho$}. 
Note that 
$\wp_{(\rho)} \int f(x) dx$ need not necessarily to converge as $\rho \rightarrow 0$. 
For neatness, we shall  write it just  $\wp \int f(x) dx$ from now on.

We now show the following integral formula, needed in our analysis:
For $0< \s <1$, 
\Beq
\wp \int_0^1 dx\  \frac{1-x}{(x^2 - \s^2)^2} 
 = \frac{1}{8\s^3}\ln\left( \frac{1+\s}{1-\s} \right)^2 +  
 \frac{1- \s}{ 2\s^2 \rho} + O(\rho) \ \ . 
\label{eq:useful}
\Eeq
To show this formula, set $f(z):= \frac{1-z}{(z^2 - \s^2)^2} $. 
Circumventing $z=\s$ from above, we enclose the 
contour in the complex plane to get
\[
\wp \int_0^1 + \int_{{\curvearrowright}_\rho (\s)}+ \int_1^{1+i R} + \int_{1+iR}^{iR} + \int_{iR}^0 =0 \ \ . 
 \]
Here the path ${\curvearrowright}_\rho (\s)$ is the same as  in {\it Appendix} \ref{app:1}; 
the paths for $\int_1^{1+i R}$ and $\int_{i R}^0$ run parallel to the imaginary axis, while 
the path for $\int_{1+iR}^{iR}$ runs parallel to the real axis. Taking the limit $R \longrightarrow \infty$, 
then, the real part of the above equation yields 
\[
\wp \int_0^1= \Re \int_0^{i\infty} -  \Re \int_1^{1+i \infty}+ \  {\cal I}(\s, 1:2)\ \ , 
\]
where the last term is given by 
Eq.(\ref{eq:singint1}) with $\xi=1$. 
It is now straightforward to compute the first two integrals on the R.H.S., yielding 
 the formula Eq.(\ref{eq:useful}). 
By the similar arguments as in {\it Appendix} \ref{app:1}, the same formula  Eq.(\ref{eq:useful}) 
is obtained 
even though we choose a contour circumventing $z=\s$ from below.

\vskip .5cm


\begin{thebibliography}{99}
\bibitem{WuFord}
Chun-Hsien Wu, Chung-I Kuo and L. H. Ford, Phys. Rev. A {\bf 65}, 062102 (2002).
\bibitem{YuFord}
H. Yu and L. H. Ford, Phys. Rev. {\bf D70}, 065009 (2004).
\bibitem{JS}
M. T. Jaekel and S. Reynaud, Quantum Opt. {\bf 4}, 39 (1992);
J. Phys. I (France) {\bf 2}, 149 (1992); {\bf 3}, 1 (1993); {\bf 3}, 339 (1993) 
\bibitem{BroMac}
L. S. Brown and G. J. Maclay, Phys. Rev. {\bf 184}, 1272 (1969).
\bibitem{DaviesDavies}
K. T. R. Davies and R. W. Davies, Can. J. Phys. {\bf 67}, 759 (1989).
\end{thebibliography}
\end{document}